\documentstyle[pre,aps,graphicx]{revtex}

\begin{document}


\title{Kinetic pathways of multi-phase surfactant systems}

\author{M. Buchanan$^\dagger$, L. Starrs$^\dagger$, S.U. Egelhaaf 
and M.E. Cates}

\address{
Department of Physics and Astronomy, The University of Edinburgh, Mayfield
Road, Edinburgh, UK EH9 3JZ\\ 
$\dagger$ Present address: Centre de Recherche Paul-Pascal, CNRS, Avenue du
Docteur-Schweitzer, F-33600 Pessac, France
}

\date{\today}

\maketitle

\begin{abstract}

The relaxation following a temperature quench of two-phase ($L_{\alpha}$ and
$L_3$) and three-phase ($L_{\alpha}$, $L_3$ and $L_1$) samples, has been
studied in an SDS/octanol/brine system.  In the three-phase case we have
observed samples that are initially mainly sponge phase with lamellar and
micellar phase on the top and bottom of the sample respectively. Upon
decreasing temperature most of the volume of the sponge phase is replaced by
lamellar phase. During the equilibration we have observed three regimes of
behaviour within the sponge phase: (i) disruption in the $L_3$ texture, then
(ii) after the sponge phase homogenises there is a $L_{\alpha}$ nucleation
regime and finally (iii) a bizarre plume connects the $L_{\alpha}$ phase with
the $L_1$ phase. The relaxation of the two-phase sample proceeds instead in two
stages. First $L_{\alpha}$ drops nucleate in $L_3$ forming an onion `gel'
structure. Over time the $L_{\alpha}$ structure compacts while equilibrating
into a two phase $L_{\alpha}$/$L_3$ sample.  We offer possible explanations for
some of these observations in the context of a general theory for phase
kinetics in systems with one fast and one slow composition variable.

\end{abstract}



\section{Introduction}

The study of nonequilibrium phase behaviour in complex fluid systems provides a
new challenge. Specifically, in the case of surfactant systems, relaxation to
equilibrium has recently attracted interest~\cite{egelhaaf:kinetics,egelhaaf:vestran,morris:emulision,egelhaaf:ostrip,siegel:lamhex,vollmer:nuc,vollmer:kin}.
For example, this has been studied in phase penetration scan
experiments~\cite{virchow:myelin,laughlin:book,argen:thesis,sakurai:myelin,sakurai:growthrates,mishima:growthrates,sakurai:structure},
and in directional growth studies~\cite{sallen:growth}.  In these experiments
the evolution of a mesophase is followed in the presence of a concentration or
temperature gradient respectively. This may result in unstable interfaces which
can lead to spectacular instabilities.

The kinetics of phase behaviour has been well studied in other areas of soft
matter physics.  For example the nucleation behaviour in colloid-polymer
mixtures has been investigated experimentally and
theoretically~\cite{evans:diff,poon:free,evans:meta,evans:phases,renth:long,evans:long}.
In these systems, three-phase samples consisting of colloidal gas, liquid and
crystal, can be prepared at different
compositions. In~\cite{poon:free,renth:long}, three-phase samples were
homogenised and observed while equilibrating. By identifying the fast and slow
composition variables in the system, and then considering the evolving
free-energy landscape for the slow component, restrictions can be placed on the
kinetic pathway the system will follow as it evolves towards equilibrium.  This
free-energy landscape theory~\cite{poon:free,evans:long} promises to be a
powerful way to predict kinetic pathways, at least in systems where one
combination of the composition variables evolves more slowly than the others,
so that time scales can be separated. It is used below (Section~\ref{sdsdisc})
to give a possible explanation for the observations in surfactant systems that
will be presented here (Section~\ref{results}).

In this paper we investigate the temporal evolution of a two-phase system,
comprising sponge ($L_3$) and lamellar ($L_{\alpha}$) phase, and a three-phase
system, comprising $L_3$, $L_{\alpha}$ and micellar ($L_1$) phase, in samples
that have been subjected to a temperature quench.  Using dark-field {\it
macroscopy} (explained below) we have observed a number of interesting
nonequilibrium features that occur as the sample progresses to its new
equilibrium state. In the two-phase sample we have observed dense nucleation of
lamellar phase droplets within $L_3$ which appear to form a `gel' of droplets
that compacts over time. In addition, we present microscopic images of the
`gel' formation.  In three-phase samples we observed three regimes of
behaviour.  Initially the sponge phase forms an {\it emulsion} which proceeds
to a homogeneous sponge phase. In the second regime nucleation of lamellar
phase in the sponge phase occurs close to the $L_3$/$L_1$ interface. The
growing droplets rise through the $L_3$ phase and arrive at the
$L_3$/$L_{\alpha}$ interface, building up the $L_{\alpha}$ phase over
time. Finally a {\it plume} pushes through the sponge phase connecting the
$L_1$ and $L_{\alpha}$. These results are described in Section~\ref{results}
and discussed in Section~\ref{sdsdisc}. Section~\ref{conc} gives a brief
conclusion.

\section{Experimental Section}
\label{expsec}

	\subsection{Materials}

SDS and octanol (both $99\%$ purity) were supplied by Sigma-Aldrich
and used without further purification.  Samples were prepared by
dissolving SDS and octanol in brine ($20$g/l of NaCl in distilled
water) and then tumbled for 2 days and subsequently allowed to
equilibrate at a higher temperature of $40^{\circ}\textrm{C}$. Experiments were
performed after approximately two weeks, when the samples had reached
equilibrium.

	\subsection{Dark-field macroscopy}

Compared to a direct observation, dark-field macroscopy allows us to
obtain better contrast from regions in the sample that scatter light
only weakly~\cite{starrs:phd,poon:gels}. The scattering of light is
caused by refractive index inhomogeneities in the sample which, {\it
e.g.}, result from regions of different concentration. The term {\it
macroscopy} is introduced since the sample is larger than its image,
which matches the size of the CCD chip, and therefore the optical
setup is designed to {\it reduce} the image size. We have used
dark-field macroscopy to observe samples whose cell dimensions are
$4\textrm{cm} \times 1\textrm{cm} \times 1\textrm{cm} $.

The general layout of the dark-field macroscope is similar to a dark-field
microscope and shown in figure~\ref{darkf}.  
A white light source and a collimator are used to produce parallel light to
illuminate the sample which is placed in the object plane. The unscattered
light is focused into the back focal plane of the objective. A beam stop is
placed in this plane to black the unscattered light. In the case of no
scattering the image is completely dark, hence the name dark-field. If there is
scattered light, it is collimated by the objective and focused onto the CCD
chip.  Therefore, whenever there are refractive index differences in the sample
these will appear as bright spots on a dark background.  All samples observed
are temperature controlled in a heat bath.

	\subsection{Microscopy}

All microscopy observations were made using an Olympus BX50 microscope
in bright-field mode between crossed polarisers with long working
distance objectives. The evolution of the samples were recorded using
a CCD camera and time lapse video recorder together with a
framegrabber PC. The contrast of the presented images has been
enhanced by thresholding. All microscopy experiments were done in a
Linkam LTS hot-stage which ensured a controlled and stable
temperature.

\section{Results}
\label{results}

We have studied two phase ($L_{\alpha}$,$L_3$) and three-phase
($L_{\alpha}$,$L_3$,$L_1$) samples upon a large temperature change. Initially
the samples are in equilibrium at $\textrm{T} = 40^{\circ}\textrm{C}$ where
they have been observed for one week and showed no further change.  The sample
is then placed in the dark-field macroscope which is kept at a temperature of
$20^{\circ}\textrm{C}$. Its behaviour upon this temperature quench is then
followed. We have studied samples with different bilayer (SDS + octanol)
concentrations and octanol/SDS ratios. We will first present the typical
behaviour of two phase and three phase samples, before we describe the
dependence on octanol/SDS ratio.

\subsection{Two-phase samples}				
\label{2p-quench}

\subsubsection{Macroscopic observations}

Temperature quench experiments on a two-phase sample ($L_{\alpha}$,$L_3$) were
performed for an octanol/SDS ratio of 1.36 and a bilayer concentration of
$30\%$.  At this composition and at a temperature of $40^{\circ}\textrm{C}$ the
sample has two phases, $L_{\alpha}$ at the top and $L_3$ on the bottom, where
the $L_3$ phase is in abundance ($\sim 95\%$).  At lower temperature
($\textrm{T} = 20^{\circ}\textrm{C}$) most of the volume ($\sim 70\%$) of the
sample is present as $L_{\alpha}$ phase which replaces $L_3$ phase.

When the sample is quenched from $\textrm{T} = 40^{\circ}\textrm{C}$ to
$20^{\circ}\textrm{C}$, we have observed two regimes of behaviour.  Almost
immediately, ($L_3$ + $L_{\alpha}$) nucleates in the $L_3$ phase and appears to
completely fill the sample (figure~\ref{twophase}A).  Over time this mixed
sponge/lamellar phase decreases in volume and the 'pure' sponge phase now
reforms at the bottom of the sample (figure~\ref{twophase}B). As the volume of
($L_3$ + $L_{\alpha}$) decreases there is a noticeable change in its
texture. The lower part is rough in texture and separated by an interface from
the top part which has a smooth texture (figure~\ref{twophase}C).  The rough
texture appears to be a gel of $L_{\alpha}$ droplets in $L_3$; this is
confirmed by microscopy observations (see below).  Over time the smooth part
grows as the rough part shrinks (figure~\ref{twophase}C-D). The final state of
the sample (not shown) contains only a smooth lamellar phase in peaceful
coexistence with the sponge phase where the two phases occupy comparable
volumes of the sample.

The same generic behaviour is observed in samples with octanol/SDS ratios up to
1.44 at 30$\%$ bilayer concentration.  In figure~\ref{gelprof1} the relaxation
to equilibrium observed with these samples is schematically presented.

\subsubsection{Microscopic observations}

To obtain more detailed information, part of the sponge phase was extracted
from a two-phase sample. It was subsequently observed with a microscope during
a temperature ramp from $\textrm{T} = 40^{\circ}\textrm{C}$ to $20^{\circ}\textrm{C}$. Droplets of $L_{\alpha}$ phase,
which have the birefringent `Maltese cross' signature of multilamellar vesicles
or onions, are observed to nucleate homogeneously in the $L_3$ phase. These
droplets continue to grow to $10-50 \mu \textrm{m}$ over the next 60 minutes
(figures~\ref{onigel}A-C). However, after 30 minutes the onions start to move
towards each other forming an onion `gel' (figure~\ref{onigel}D). This
structure coarsens over time leaving voids of pure sponge phase
(figure~\ref{onigel}E-F).  As more onions form in the $L_3$ phase they
coalesce. The heterogeneity of the gel is presumably responsible for the rough
texture reported above.

Figure~\ref{spongeoni} shows a close-up of two droplets, with Maltese crosses
visible, when observed through partially crossed polars. In this case the two
droplets have partially merged.  On closer inspection these lamellar droplets
have a distinctive pattern of focal conic defects that form within the onion
(figure~\ref{epitax}A). These defects in the onion are presumably due to an
epitaxial tilt of the bilayers at the surface of the droplet, which is
surrounded by sponge phase. This has also been observed at the interface
between a homeotropic lamellar and sponge phase~\cite{blanc:curv}.  The
epitaxial tilt is required to match the repeat distance in the lamellar phase
onto the characteristic bilayer spacing in the sponge (figure~\ref{epitax}B).
This observation suggest that, at the observed temperature and time scales, the
internal structure of each onion is capable of reorganizing to minimize the
interfacial energy of the onion.

	\subsection{Three-phase samples} 		\label{3p-quench}

Samples within the three-phase region have been observed with $5\%$ and $10\%$
bilayer concentration over a range of octanol/SDS ratio $1.22$ to $1.32$. At
these compositions and at $\textrm{T} = 40^{\circ}\textrm{C}$, all samples
consist of three phases: $L_{\alpha}$ at the top, $L_3$ in the middle and $L_1$
on the bottom, where the $L_3$ phase has the largest volume.  At a lower
temperature of $20^{\circ}\textrm{C}$ most of the volume is occupied by
$L_{\alpha}$ phase with only a small volume of $L_3$ phase remaining. The $L_1$
phase is observed to have a slightly larger volume at $\textrm{T} =
20^{\circ}\textrm{C}$ than at $40^{\circ}\textrm{C}$. After a temperature
quench from $\textrm{T} = 40^{\circ}\textrm{C}$ to $20^{\circ}\textrm{C}$ the
relaxation to equilibrium is found to proceed through three regimes. They are
described below for a sample that has a bilayer concentration of $5\%$ and an
octanol/SDS ratio of $1.22$. Other three-phase samples behave in a similar
fashion, although the time constants were found to depend on the actual
composition.

        \subsubsection{Early stages}

After the temperature quench, the $L_3$ phase becomes turbid as inhomogeneities
form within it (figure~\ref{TquenchI}A).  This texture appears to be an
emulsion of $L_1$ droplets separated by $L_3$ or vice versa.  (see
Section~\ref{sdsdisc} for a discussion).  The texture is first seen close to
the $L_1$/$L_3$ interface, and then spreads up to the $L_3$/$L_{\alpha}$
interface while large convection currents are also observed within the $L_3$
phase. 

In addition some large droplets of $L_1$ fall rapidly from the
$L_{\alpha}$ phase through the inhomogeneous $L_3$ phase into the $L_1$ phase
(figure~\ref{micdrops}).  
These droplets are initially of a diameter of about
$90 \mu \textrm{m}$ and fall with speed $\sim 1000 \mu \textrm{m}
\textrm{s}^{-1}$. Within 3 hours their size decreases to $\sim 14 \mu
\textrm{m}$ and their speed is reduced to $\sim 15
\mu \textrm{m} \textrm{s}^{-1}$.  About one hour later the $L_3$ phase becomes
homogeneous once again, by a clearing process which starts close to
the $L_3$/$L_{\alpha}$ interface. A schematic summary of these
features is given in figure~\ref{timeprof}. These features are typical
of samples at $5-10\%$ bilayer composition with octanol/SDS ratios
$1.22-1.32$.

        \subsubsection{Lamellar nucleation and growth} 	\label{sec:nucgro} 

After about one hour, droplets of $L_{\alpha}$ phase of approximately $10-20
\mu \textrm{m}$ in size begin to form in the $L_3$ phase.  The number of
$L_{\alpha}$ drops observed increases over the next 30 minutes.  While droplets
are observed to nucleate throughout the $L_3$ phase, this mainly happens near
the $L_1$/$L_3$ interface (figure~\ref{TquenchII}).  The droplets move up
towards the $L_{\alpha}$ phase with a speed of about 4$\mu \textrm{m}
\textrm{s}^{-1}$, gaining size as they do so. Over a period of $1-2$ days they
build the $L_{\alpha}$ phase. The quantity of the $L_{\alpha}$ phase steadily
increases, however after 12 hours the number of droplets within $L_3$ is
observed to have decreased. Eventually after about 36 hours further droplets
are not observed to form despite the continuing growth of the lamellar
phase. These findings are summarized schematically in Fig.~\ref{nucprof}.

At larger bilayer concentrations ($30-40\%$) we have observed a much
larger density of lamellar droplets in the $L_3$ phase. At a much
larger octanol/SDS ratio of 1.35 and 1.41 the droplets that form are
very dense. Furthermore, by increasing the octanol/SDS ratio, we have
observed a noticeable change in the dynamics. At higher octanol/SDS
ratio we observe an increase in the rate of formation of the
$L_{\alpha}$ droplets and the speed of the droplets as they move
towards the $L_{\alpha}$ phase.

        \subsubsection{The plume}

After 36 hours the $L_{\alpha}$ droplets are no longer observed to
nucleate in the $L_3$ phase. However, the $L_{\alpha}$ phase still
continues to grow.  Eventually at some later time a plume grows from
the $L_1$ phase and extends up to the $L_{\alpha}$ phase
(figure~\ref{plume}).

Prior to this, in the $L_1$ phase small droplets of $L_{\alpha}$ phase are
observed to move up towards the $L_3$/$L_1$ interface and eventually push their
way through the $L_3$ phase by forming the plume.  The plume connects the $L_1$
phase to the $L_{\alpha}$ phase cutting straight through the $L_3$ phase. It is
about $50-100 \mu \textrm{m}$ in diameter. On the plume near the $L_1$/$L_3$
interface a large bulbous piece with a diameter of about $100 \mu \textrm{m}$
can be observed. We followed the movement of this over the next 12
hours. Initially the bulbous piece rises up toward the $L_{\alpha}$/$L_3$
interface (figure~\ref{plume}B-C). Then it begins to move back towards the
other interface (figure~\ref{plume}C-E). Eventually the plume detaches from the
$L_3$/$L_1$ interface and the whole plume is `pulled' up into the $L_{\alpha}$
phase (figure~\ref{plume}E-G) while the bulbous bit remains at the loose end of
the plume. The only way to describe the motion of the plume is that it behaves
in a `lava-lamp' like fashion.

The occurrence of the plume is a reproducible effect and has been seen
in many separate experiments. However, in the range of samples we have
observed we only find plumes when there is $L_1$ phase
present. Depending on the composition we can even observe multiple
plumes forming. The plumes are observed late in the lamellar
nucleation regime. However, at larger octanol/SDS ratios plumes form
earlier in the $L_{\alpha}$ nucleation regime. In addition, we have
observed streaks in the $L_3$ phase at the $L_{\alpha}$ interface,
which persist for the remaining duration of the growth of
$L_{\alpha}$. They are found during and after the formation of the
plume.

\subsubsection{Octanol/SDS dependence}

As already mentioned in section~\ref{sec:nucgro}, a higher octanol/SDS ratio
also results in an increase in the rate of formation of the
$L_{\alpha}$ droplets and the speed of the droplets as they move
towards the $L_{\alpha}$ phase. In addition, the density of droplets
is increased upon an increase in the octanol/SDS ratio.  By varying
the octanol content in the bilayer we influence the time scale of each
regime that is observed. The initial regime, where we have observed
the transient formation of a sponge emulsion, appears to persist for a
shorter period of time as we increase the octanol/SDS ratio.

This is also the case for the second regime of lamellar nucleation and
growth. Furthermore the nucleation regime starts earlier for samples
with higher octanol content so that these regimes overlap. This is
also the case for the observation of the plume which can be observed
earlier at higher octanol/SDS ratio. Figure~\ref{regdiag} gives a
semi-quantitative diagram of these observations.

\section{Discussion}				\label{sdsdisc}

An initially equilibrated sample that is subjected to a sudden
temperature change will relax to a new equilibrium state. In three-
and two-phase surfactant samples we have observed interesting
behaviour as the samples proceed to this final equilibrium
state. Below we discuss these observations first for the three- and
then for the two-phase samples. This will be followed by some possible
explanations for the observed behaviour.

            \subsection{Early stages}

When the three phase sample is quenched from $\textrm{T} =
40^{\circ}\textrm{C}$ to $20^{\circ}\textrm{C}$ the $L_1$ phase volume is found to increase
during the early stages.  This indicates that during the quench the majority
phase, $L_3$, increases its bilayer concentration with a concomitant reduction
in its bilayer spacing, expelling $L_1$. This would explain the formation of a
`sponge emulsion', which appears as an inhomogeneous texture
(figure~\ref{TquenchI}).  A minority of $L_1$ phase in sponge phase has
previously been observed in other experiments~\cite{strey:AOT,buchanan:AOT}.
Furthermore, the subsequent separation of $L_1$ and $L_3$ phase may lead to the
observed convection currents.  The sample begins to homogenise again as the
compositions reorganise into a more favourable state. The homogenisation of the
$L_3$ phase is found to start at the $L_{\alpha}$ phase and moves towards the
$L_1$ phase. This is not surprising, since the $L_1$ phase has to drain through
the sponge emulsion during the homogenisation process (figure~\ref{TquenchI}).

A similar decrease in bilayer spacing is also expected for the
$L_{\alpha}$ phase. Consequently the bilayer concentration will
increase and excess brine will be expelled, possibly together with a
low concentration of surfactant and cosurfactant. It seems conceivable
that this leads to the formation of droplets of $L_1$ within
$L_{\alpha}$, which are then observed to fall through the $L_3$ phase
(figure~\ref{micdrops}). As the equilibrium bilayer concentration is
approached, the $L_1$ flux is reduced. This is reflected in the
reduction of the $L_1$ droplet size and consequently leads to a
reduction in the speed of the droplets.

By expelling excess brine, both the $L_3$ and $L_{\alpha}$ phase
increase their bilayer concentration.  The decrease in undulation
repulsions between layers as a result of the lowered temperature seems
likely to be responsible for this
behaviour~\cite{helfrich:undulation}. This change in bilayer
concentration occurs during the early stages and appears to be much
faster than reorganisation within the bilayer. The slower
intra-bilayer reorganisation is thus expected to influence the
behaviour during the subsequent stages.

        \subsection{Lamellar nucleation and growth}

After the early stage described above (and sometimes overlapping with
it -- see figure~\ref{regdiag}), lamellar phase is found to nucleate
and grow within $L_3$ at the expense of sponge phase. One of the most
striking features of this process is that the nucleation of the
$L_{\alpha}$ phase in the $L_3$ phase is not homogeneous throughout
the $L_3$ phase, but is most abundant close to the $L_1$/$L_3$
interface. It seems peculiar that such nucleation should occur in the
part of the $L_3$ region {\it furthest} from the existing slab of
$L_\alpha$ phase. A possible explanation for this can be found by
analysing the way the free energy landscape changes after the
temperature jump~\cite{poon:free,evans:long}.

First let us consider a three phase sample ($L_1$,$L_3$,$L_{\alpha}$)
at equilibrium. To our knowledge there is no published phase diagram
for this system that would show the phase boundaries close to the
three-phase region. However, from our experience with these samples we
present a schematic phase diagram of this region as a function of the
volume fraction of SDS $\phi_{\rm SDS}$ and octanol $\phi_{\rm oct}$
(figure~\ref{fsphase}).

Figure~\ref{landscape6} gives a schematic free energy landscape that is
consistent with the observed phase diagram of the system. The free energy of
the sample in equilibrium is a function of the volume fractions of two
independent components, SDS and octanol.

At equilibrium the chemical potentials ($\mu_{\rm SDS,oct} = \partial
f/\partial \phi_{\rm SDS,oct}$) of each component must be equal in all of the
coexisting phases, $\mu_{\rm SDS,oct}^{L_1} = \mu_{\rm SDS,oct}^{L_3} =
\mu_{\rm SDS,oct}^{L_\alpha}$. The same also applies to the osmotic pressure of
each phase.  This means that the tangents (shaded plane, Fig.~\ref{landscape6})
to the free energy `wells' at the equilibrium compositions will all be the
same.  The points of tangency will correspond to the corners of the three-phase
triangle on the phase diagram (figure~\ref{fsphase}).

Note that this analysis is invariant under arbitrary linear
combination of the composition variables. Here we will choose linear
combinations $\phi_{\rm f},\phi_{\rm s}$ as indicated in
figure~\ref{fsphase}. (These are related to the volume fraction of
bilayer and the composition of the bilayer, respectively, as we
discuss below.) With these as the densities, the phase diagram looks
like figure~\ref{landscape4}.

Now we ask: How does a three-phase sample (marked in
figure~\ref{landscape4} by `$\bullet$') change upon a temperature
quench? Suppose at a given temperature its composition is close to the
$L_3$ corner of the three-phase triangle (gray area in
figure~\ref{landscape4}). It thus mainly consists of $L_3$, but also
contains small amounts of $L_{\alpha}$ and $L_1$. Upon a change in
temperature the free energy landscape rearranges such that the free
energy minima move to new positions. This changes the position of the
three phase triangle (dashed lines in
figure~\ref{landscape4}). Although the chemical composition of the
sample remains constant, its relative position in the three-phase
region is changed. In the case shown, it is now located much closer to
the $L_{\alpha}$ corner and also moved towards the $L_1$ corner. The
sample is thus forced to change its compositions locally and nucleate
$L_{\alpha}$ and $L_1$ at the expense of $L_3$. This is what happens
in our three-phase samples when we quench from
$40^{\circ}\textrm{C}$ to $20^{\circ}\textrm{C}$.

Following reference~\cite{poon:free}, a quasi-equilibrium free energy
landscape can be constructed after the temperature quench if fast and
slow composition variables can be identified. Application of this {\it
free energy landscape} analysis to our system depends on the relation
between the fast and slow components ($\phi_{\rm f},\phi_{\rm s}$) and
the volume fractions of octanol ($\phi_{\rm oct}$) and SDS ($\phi_{\rm
SDS}$).  Our observations indicate that changes in bilayer
concentration occur during the early stages and can thus be identified
with the fast component. In contrast, changes in the organisation of
the bilayer such as the transition from $L_3$ to $L_{\alpha}$ are
found to happen later. From the phase diagram (figure~\ref{fsphase})
it is clear that this $L_3$-$L_{\alpha}$ transition must involve
changes in the bilayer composition (at nearly constant bilayer
fraction).  This is also supported by our separate observations of
contact experiments between lamellar phase and water, where myelin
patterns form~\cite{buchanan:AOT,buchanan:myelins}.

Immediately after contact the myelin formation is always observed
before the formation of any other intermediate phase, such as $L_3$
phase.  Thus the time for bilayers to swell to the miscibility gap is
faster than the formation of sponge phase which requires local
reorganisation in the bilayer structure, and also a change in its
composition.

We therefore assume that, starting from a given initial state, changes
in volume fraction of the fast component $\phi_{\rm f}$ correspond to
changes in total bilayer concentration at fixed SDS/octanol
ratio. (This is expected to be a fast, `collective diffusion' mode.)
We also assume that the slow component corresponds to changes in
SDS/octanol ratio at roughly fixed bilayer volume fraction. This
second assumption is physically reasonable, but less obvious:
formally, the fast and slow directions are eigenvectors of a diffusion
matrix, which need not be symmetric, and they need not be at
rightangles. However, if we make both assumptions, the fast and slow
direction correspond to radial and tangential directions in
figure~\ref{fsphase}, which is probably a good enough approximation
for the present discussion. For a state near the $L_3$ corner
initially, these are as indicated in the figure.

For the fast component, the chemical potential $\mu_{\rm f} = \frac{\partial
f}{\partial \phi_{\rm f}}$ will become uniform rapidly.  Essentially, we move
to an equilibrium state through a sequence of metastable states, for which the
chemical potential of the fast component $\mu_{\rm f}$ is uniform. However, the
value of $\mu_{\rm f}$ is evolving in time and depends on the (nonuniform)
evolution of the slow variable $\phi_{\rm s}$.  Instead of drawing the free
energy landscape as a function of two components (figure~\ref{landscape6}) we
now concentrate on the slow variable $\phi_{\rm s}$ only.  (This defines a free
energy $f(\phi_{\rm s},\mu_{\rm f})$, as a function of $\phi_{\rm s}$ in a
semi-grand ensemble, under conditions of uniform, but time dependent, $\mu_{\rm
f}$.)  Figure~\ref{phspace}A shows the initial (equilibrium) free energy
landscape at high temperature as a function of the slow composition
variable. After a temperature quench the three minima do not have a common
tangent (figure~\ref{phspace}B) since $\mu_{\rm f}$ has not yet relaxed to its
equilibrium value and the initial condition now represents a metastable
state. In this diagram we may assume that the free energy curve of $L_3$ phase
lies higher than the common tangent of the other two so as to favour producing
more $L_1$ and $L_{\alpha}$ phase. (If the free energy minimum of $L_3$ phase
is below the tangent of the other two then no nucleation will occur.  In fact,
this would be the case where the change in phase diagram is in the opposite
direction to the one shown in figure~\ref{landscape4}.)  Once the sample
reaches equilibrium at the lower temperature, the three minima again have a
common tangent (a situation similar to figure~\ref{phspace}A is recovered).

The local composition of the phases near their interfaces with one
another can be determined from the {\it metastable common tangents} to
the free energy landscape (figure~\ref{phspace}B); the compositions at
each end of a tangent have the same chemical potential $\mu_{\rm s}$
of the slow component which is the condition for {\it local}
equilibrium of the interface.  From these tangents, the local
composition of the $L_3$ phase near to its interfaces can be
determined. The $L_3$ phase will reorganise to have two different
compositions, $\phi_{\rm s}=\phi_1$ near the $L_{\alpha}$/$L_3$
(upper) interface and $\phi_{\rm s}=\phi_2$ near the $L_3$/$L_1$
(lower) interface. The $L_3$ phase thus has a higher bilayer volume
fraction nearer the $L_3$/$L_1$ interface. The resulting composition
profile as a function of sample height $z$ is shown in
figure~\ref{phipro}.  Since the nucleation of the phases $L_{\alpha}$
and $L_1$ will be induced by composition fluctuations in $L_3$ phase,
$L_{\alpha}$ is more likely to nucleate close to the $L_3$/$L_1$
interface with its higher volume fraction.  The difference in
concentrations at the two interfaces requires a concentration gradient
across the $L_3$ phase. Since $\phi_1 < \phi_2$ then the upper part of
the phase has higher mass density than the lower part. (This could
cause the system to become unstable under gravity.)
The concentration gradient as well as gravity favours driving the $L_{\alpha}$
droplets up towards the $L_{\alpha}$ phase.  

Note that these features directly depend on the presence of the $L_1$
phase which fixes $\phi_{\rm s} = \phi_2$ at the bottom of the sponge
phase. Also, the inversion of the concentration profile predicted by
this model follows directly from the fact that the homogeneous $L_3$
region in the initial state is metastable: its common tangents to
$L_1$ and $L_{\alpha}$ cross each other (figure~\ref{phspace}B).

        \subsection{The plume}

The formation of a plume which extends from the $L_1$/$L_3$ interface
into the $L_{\alpha}$ phase appears to result from the nucleation of
lamellar phase in the $L_1$ phase. The formed lamellar phase might
build up at the $L_1$/$L_3$ interface, because the density difference
between $L_{\alpha}$ and $L_3$ is very small and the surface tension
between $L_{\alpha}$ and $L_3$ is large.  Eventually the $L_{\alpha}$
phase formed in the $L_1$ phase pushes through the sponge phase in a
`lava-lamp' like fashion.

            \subsection{Comparison of two- and three-phase samples}

The absence of the $L_1$ phase leads to a clearly different relaxation
behaviour of the two-phase samples, although for both kinds of samples
the main change upon a temperature quench from $\textrm{T} =
40^{\circ}\textrm{C}$ to $20^{\circ}\textrm{C}$ is
the formation of $L_{\alpha}$ at the expense of $L_3$ phase. There is
no indication for the formation of a (metastable) $L_1$ phase within
the $L_{\alpha}$ or $L_3$ phase during the early stages as observed
for the three-phase samples. This could, however, be due to the large
octanol/SDS ratio of the two-phase samples, which leads to an overlap
of the early stages and the lamellar nucleation in the case of the
three phase sample (figure~\ref{regdiag}) and could mask the
(intermediate) formation of $L_1$.

In the case of the two-phase samples, the nucleation of $L_{\alpha}$
occurs homogeneously throughout all of the $L_3$ phase. Since only one
interface ($L_{\alpha}$/$L_3$) is present, the $L_3$ phase cannot
develop a density inversion which could influence the local nucleation
rate. The quantity of nucleating $L_{\alpha}$ is large due to the
relatively high bilayer concentration ($30\%$). The nucleated
$L_{\alpha}$ droplets (onions) appear to form a dense network in the
$L_3$ phase (figure~\ref{onigel}), similar to a colloidal gel
structure.  This is followed by a compaction process, which could be
viewed as the drainage and collapse of a `gel' of $L_{\alpha}$
droplets in the $L_3$ phase. (For analogous phenomena in colloids
see~\cite{poon:gels,starrs:phd}.)  Due to the lower mass density of
the $L_{\alpha}$ phase, the onion `gel' collapses towards the upper
phase boundary of the $L_3$ phase. This results in the observed smooth
texture above the still uncompacted onion gel in the $L_3$ phase with
its rough texture (figures~\ref{twophase} and~\ref{gelprof1}). This is
reminiscent of the creaming of emulsions. The final equilibrium state
of the two-phase samples consists of a completely compacted lamellar
phase and `pure' $L_3$ phase.  Since there is no $L_1$ phase present,
it is no surprise that the plume has not been observed in two-phase
samples.

			\subsection{Other features}

We have performed a small number of experiments in which the
temperature quench was reversed before the reorganization of the
phases was complete.  During an {\it increase} in temperature, we have
observed {\it smectic bubbles} in the $L_3$ phase.  These are
$L_{\alpha}$ films which contain focal conic defects
(figure~\ref{onispon}).  Each bubble once contained a lamellar onion
droplet suspended in the $L_3$ phase. On raising the temperature, the
central region melts and we observe changes in the birefringence of
the bubble.

Eventually the bubble rapidly shrinks down into a small lamellar
droplet (figure~\ref{burst}). In some cases we observe the bubble to
rapidly collapse and then completely vanish after it bursts. The
kinetic behaviour of such processes remains to be fully investigated.

\section{Conclusion}\label{conc}

Temperature quench experiments in surfactant systems have led to a
variety of interesting observations. In the initial stages after the
temperature quench, disruption of the $L_3$ phase occurs and $L_1$
droplets are found to move from the $L_{\alpha}$ through the $L_3$ to
the $L_1$ phase.  This is attributed to a sudden reduction in the
bilayer concentration which causes the $L_3$ and $L_{\alpha}$ phase to
expel solvent forming a `sponge emulsion', and a suspension of $L_1$
droplets, respectively. This is only observed in the first few hours
after the quench. After this rapid equilibration of the bilayer
concentration, we observe the formation of lamellar droplets in the
sponge phase close to the $L_1$/$L_3$ interface.  This observation can
be explained in terms of the evolution through a free energy
landscape.  As the system proceeds to equilibrium it goes through a
series of metastable states. Under such conditions the bilayer
concentration at the $L_1$/$L_3$ interface can be larger than at the
$L_{\alpha}$/$L_3$ interface. In this case the sponge phase is more
likely to nucleate lamellar phase droplets close to the
micellar/sponge interface. However until we can directly probe the
density of the sponge phase as a function of sample height during this
process this explanation remains speculative, and the inverted density
profile of figure~\ref{phipro} remains a conjecture.

At a later stage we observe a plume to form at the $L_1$/$L_3$
interface which connects up to the $L_{\alpha}$ phase. This may be due
to the nucleation of lamellar phase droplets in the micellar phase,
$L_1$ which get `stuck' when they reach the sponge phase interface. If
the surface tension of the $L_{\alpha}$/$L_3$ interface is large then
the lamellar phase will build up until there is enough to break
through. The break through is driven by gravity as in a lava-lamp.

In two-phase samples which were prepared at a larger bilayer concentration we
have observed different features. The absence of the third phase ($L_1$)
means we do not observe the nucleation at the bottom of the cell or the
formation of the plume. This gives weight to the free-energy landscape
explanation of the nucleation kinetics.

\section*{Acknowledgements}
We are grateful to P. Garrett, D. Roux, J. Leng, G. Tiddy, J. Walsh and
P. Warren for useful discussions. LS thanks the Institute of Food Research for
a CASE award.  MB thanks Unilever PLC and EPSRC (UK) for a CASE award.
Work funded in part under EPSRC Grant GR/K59606.


\begin{figure}
\centering
\caption{Schematic diagram of dark-field macroscopy.} 
\label{darkf}
\end{figure}

\begin{figure}
\centering
\caption{Temperature quench from $\textrm{T} = 40^{\circ}\textrm{C}$ to
$20^{\circ}\textrm{C}$ of a two phase sample. (A) Immediately after the
temperature quench a mixture of $L_3$ and $L_{\alpha}$ phase appears to
completely fill the sample. (B) Its volume begins to decrease and 'pure' sponge
phase is observed at the bottom of the cell. (C-D) Two textures of lamellar
phase are observed: rough (middle) and smooth (top). The smooth part increases
in volume as the rough part decrease over time. (Times displayed in minutes.)}
\label{twophase}
\end{figure}

\begin{figure}
\centering
\caption{Schematic temporal profile of lamellar ($L_{\alpha}$) and sponge
($L_3$) phases in a two phase sample. The $L_{\alpha}$/($L_{\alpha}$ + $L_3$)
interface (filled squares) is determined by the difference in smooth and rough
texture. Likewise the ($L_{\alpha}$ + $L_3$)/$L_3$ interface is marked by
`$\bullet$'.}
\label{gelprof1}
\end{figure} 

\begin{figure}
\centering
\caption{Micrographs of the nucleation and gelation of droplets of $L_{\alpha}$
(onions) in the $L_3$ phase.  (A-C) Onions nucleate homogeneously in the $L_3$
phase and then proceed to grow.  (D-F) Subsequently onions move towards each
other forming a ``gel'' network of onions; voids of $L_3$ phase can be clearly
observed. All images viewed through crossed polars. (Times shown in minutes.)}
\label{onigel}
\end{figure}

\begin{figure}
\centering
\caption{Onions that have nucleated in $L_3$ phase after a temperature quench
and partially coalesced. Note the distinctive pattern of focal conic defects.}
\label{spongeoni}
\end{figure}

\begin{figure}
\centering
\caption{(A) Schematic diagram of an
onion in $L_3$ phase where focal conic defects form in the onion to satisfy an
epitaxial tilt at the $L_{\alpha}$/$L_3$ interface, as shown in (B).}
\label{epitax}
\end{figure}

\begin{figure}
\centering
\caption{Initial
stages of a temperature quench on a three-phase sample. At the top $L_{\alpha}$
phase (appears dark here), $L_3$ phase in the middle and $L_1$ phase at the
bottom (appears dark). (A) Disruption appears immediately after the quench
(B-C) which clears over one hour period. Times shown in minutes.}
\label{TquenchI}
\end{figure} 

\begin{figure}
\centering
\caption{Droplets of $L_1$ phase (indicated by arrows) nucleate in the
$L_{\alpha}$ phase and fall through the $L_3$ phase. The $L_{\alpha}$ phase is
on top of the $L_3$ phase and not shown in this closeup. (Times shown in
minutes.)}
\label{micdrops}
\end{figure} 

\begin{figure}
\centering
\caption{Temporal evolution of the
sample, in particular the sponge emulsion, during the early
stages. This sample contains a higher bilayer concentration of $8\%$
than the samples in figures~\ref{TquenchI} and~\ref{micdrops}; while
the generic behaviour is the same, the observed kinetics are
slower. The sponge emulsion texture is shown as the shaded region in
the diagram. The error bar indicates the width of the interface.}
\label{timeprof}
\end{figure}

\begin{figure}
\centering
\caption{Birefringent droplets form in the sponge phase, $L_3$ (middle) as the
lamellar phase, $L_{\alpha}$ (top) grows. (A-B) The initial sample settled in a
homogeneous phase. (C-F) The $L_{\alpha}$ droplets nucleate in abundance but
mainly near the lower ($L_1$/$L_3$) interface. The droplets rise to the
$L_3$/$L_{\alpha}$ interface, building the $L_{\alpha}$ phase. (G-H)
$L_{\alpha}$ drop nucleation slows down and eventually stops although the
lamellar phase continues to grow. Times shown in minutes.}
\label{TquenchII}
\end{figure}

\begin{figure}
\centering
\caption{Temporal evolution of the density of lamellar phase
droplets in the sponge phase. The positions of the $L_1$/$L_3$ and
$L_3$/$L_{\alpha}$ interfaces are given ($\Box$).  Freshly nucleated droplets
({\tiny $\bullet$}) are observed to mainly nucleate near the $L_1$/$L_3$
interface. As droplets grow and move towards the $L_{\alpha}$ phase their
density ($\bullet$) appears to be fairly uniform.}
\label{nucprof}
\end{figure} 

\begin{figure}
\centering
\caption{Bizarre plume forms at the $L_1$/$L_3$ phase interface and grows up to
the lamellar phase. Times shown in hours. }
\label{plume}
\end{figure}

\begin{figure} 
\centering
\caption{Dependence of the temporal evolution of the different regimes on the
octanol/SDS ratio. The duration and onset time of these features is also
dependent on this ratio. The total bilayer concentration is $10\%$.}
\label{regdiag}
\end{figure}

\begin{figure}
\centering  
\caption{Schematic phase diagram of the SDS/octanol/brine system close to the
three-phase region consisting of lamellar ($L_{\alpha}$), sponge
($L_3$) and micellar ($L_1$) phase. The direction of increasing
bilayer volume fraction $\phi_{\rm f}$ and relative amount of octanol
in the bilayer $\phi_{\rm s}$ are also shown.}
\label{fsphase}
\end{figure}

\begin{figure} 
\centering
\caption{Free energy landscape for an equilibrated two-component system
($f(\phi_{\rm oct},\phi_{\rm SDS})$) under conditions where three phases
exist. The shaded plane represents the tangent plane.}
\label{landscape6}
\end{figure}

\begin{figure} 
\centering
\caption{Schematic phase diagram close to
the three-phase region for two different temperatures. The solid lines
and the gray area represent the three-phase region at high
temperature. Upon a decrease in temperature the three phase region
moves to the position indicated by the dashed lines. Phases denoted
are micellar ($L_1$), sponge ($L_3$) and lamellar ($L_{\alpha}$)
phase. The composition of the sample discussed in the text is
indicated by `$\bullet$'.}
\label{landscape4}
\end{figure} 

\begin{figure}
\centering
\caption{ Free energy landscape of a three phase sample (A) in equilibrium at
high temperature, (B) during relaxation to a new equilibrium after a
temperature quench}
\label{phspace}
\end{figure}

\begin{figure} 
\centering
\caption{Schematic diagram of a three-phase sample showing the nucleation of 
$L_{\alpha}$ phase in $L_3$ phase near to the $L_1$/$L_3$ interface
and its composition profile as a function of sample height $z$ as
deduced from the free energy landscape as a function of the slow
component $\phi_{\rm s}$.}
\label{phipro}
\end{figure} 

\begin{figure}
\centering
\caption{After an increase in temperature smectic bubbles are observed to
form in $L_3$ phase.}
\label{onispon}
\end{figure}

\begin{figure}
\centering
\caption{One of the smectic bubbles `bursts'
and collapses rapidly to a point (over a period of about $1$s).}
\label{burst} 
\end{figure}

\end{document}